\documentclass[12pt]{iopart}
\usepackage{iopams}
\usepackage{epsfig}
\newcommand{\be}{\begin{equation}}
\newcommand{\ee}{\end{equation}}
\newcommand{\bd}{\begin{displaymath}}
\newcommand{\ed}{\end{displaymath}}
\newcommand{\BE}{\begin{eqnarray}}
\newcommand{\EE}{\end{eqnarray}}

\newcommand{\bra}{\left\langle}
\newcommand{\ket}{\right\rangle}

\newcommand{\sgn}{{\rm sgn}}
\newcommand{\erf}{{\rm erf}}
\newcommand{\id}{{\rm 1\!\!I}}

\newcommand{\bR}{\ensuremath{\mathbf{R}}}

\newcommand{\qbo}{{\mbox{\boldmath $q$}}}

\newcommand{\boldxi}{{\mbox{\boldmath $\xi$}}}
\newcommand{\boldomega}{{\mbox{\boldmath $\omega$}}}
\newcommand{\boldpsi}{{\mbox{\boldmath $\psi$}}}

\newcommand{\boldOmega}{{\mbox{\boldmath $\Omega$}}}

\newcommand{\tp}{t^\prime}
\newcommand{\td}{t^{\prime\prime}}

\begin{document}

\title[Statistical Mechanics of Dilute Batch Minority
Games]{Statistical Mechanics of Dilute Batch Minority Games with
Random External Information}

\author{Tobias Galla\dag\ddag$\|$
}

\address{\dag The Rudolf Peierls Centre for Theoretical Physics,
Department of Physics, University of Oxford, 1 Keble Road, Oxford OX1
3NP, UK}

\address{\ddag\ International Center for Theoretical Physics, Strada Costiera 11, 34014 Trieste, Italy}
\address{$\|$ Istituto Nazionale per la Fisica della Materia (INFM), Trieste-SISSA Unit, V. Beirut 2-4, 34014 Trieste, Italy}

\begin{abstract}
We study the dynamics and statics of a dilute batch minority
game with random external information. We focus on the case in which
the number of connections per agent is infinite in the thermodynamic
limit. The dynamical scenario of ergodicity breaking in this model is
different from the phase transition in the standard minority game and
is characterised by the onset of long-term memory at finite integrated
response. We demonstrate that finite memory appears at the AT-line obtained
from the corresponding replica calculation, and compare the behaviour
of the dilute model with the minority game with market impact
correction, which is known to exhibit similar features.
\end{abstract}


\ead{\tt galla@ictp.trieste.it}

\section{Introduction}
The minority game (MG) has attracted much interest in the statistical
physics community over the past years. It models a simple market in
which the trading agents at each time-step have to make a decision
whether to buy or sell based on publicly available information, such
as the past market-history, political circumstances, the weather
forcast etc. Each agent aims to make profit by making the opposite
choice to the majority of agents. The interaction between the agents
is indirect, the players cannot identify the actions of individual
other agents, but react to the total cumulative outcome of all actions
in the market. In order to take their trading decisions each agent
holds a pool of strategies, which can be understood as look-up tables,
mapping the value of the publicly available information onto a trading
decision. The identification of the best strategy is based on virtual
scores the agents allocate to each of their strategies in order to
keep track of their performance. At each round of the game each agent
updates the score of each of his or her strategies, increasing the
scores of strategies which would have predicted the correct minority
decision. Exhaustive reviews on MGs can be found in \cite{Chalweb, Moro04}.

Although the update rules seem simple, the MG displays remarkably
complex features. The behaviour of the fluctuations of the total bid,
the so-called volatility is non-trivial \cite{ChalZhan97, Cava99,
SaviManuRiol99,CavaGarrGiarSher99, ChalMarsZecc00} and a
phase-transition separating an ergodic and a non-ergodic phase was
identified \cite{ChalMarsZhan00,
ChalMarsZecc00,MarsChalZecc00,GarrMoroSher00,HeimCool01}.  The main
control parameter in MGs is the ratio $\alpha=P/N$ of the number $P$
of possible values for the public information over the number $N$ of
players.

From the point of view of statistical mechanics, the MG is an extremely
interesting model. It contains the ingredients of a disordered
mean-field system and can be approached with the techniques of
spin-glass physics, if the thermodynamic limit of an infinite number
of players is considered.

Although the model lacks detailed balance, it was shown that the
stationary state of the adaptive dynamics can be captured by the
minima of a disordered Hamiltonian $H$ for $\alpha>\alpha_c=0.3374...$
\cite{MC01}. Its ground states were calculated using the replica method
\cite{ChalMarsZecc00,MarsChalZecc00,ChalMarsZhan00}, and the location
of the ergodic/non-ergodic phase transition, $\alpha_c$, was
identified analytically as the point where the static susceptibility
obtained within a replica-symmetric ansatz becomes infinite. While
replica symmetry is not broken even in the non-ergodic phase of the
standard MG, later studies of MGs in which the agents correct for
their own contribution to the total bid then revealed the presence of
phases with replica symmetry breaking \cite{DeMaMars01}.

Dynamical studies of the MG have been conducted in
\cite{HeimCool01,CoolHeimSher01} using the exact generating functional
technique of \cite{DeDom78}, and enabled a full understanding of the
game in the ergodic phase. The phase transition in the standard MG was
identified as the point $\alpha=\alpha_c$ where the assumption that
the dynamical susceptibility (or integrated response function) be
finite breaks down. A review of dynamical analyses of MGs can be found
in \cite{Coolenreview}. It was also shown that MGs with market-impact
correction exhibit a different type of phase-transition
\cite{HeimDeMa01}. While the integrated response remains finite in
such models, long-term memory was identified as the source of
ergodicity breaking, and it was seen that the onset of long-term
memory coincides with the AT-line, at which the replica-symmetric
solution becomes unstable \cite{HeimDeMa01, DeMaMars01}.

On the technical level the MG is closely related to the Hopfield model
of neural networks \cite{Hopf82}, the update rules of the MG are similar to
`anti-Hebbian' learning rules of neurons. While the quenched patterns
in the Hopfield model form attractors of the learning dynamics, the
agents in the MG try to avoid such fixed points. Up to an additional
random field the above Hamiltonian of the MG is therefore
essentially identical to the Hamiltonian of the Hopfield model with
negative sign. The energy landscape of a suitably defined majority
game coincides with that of the Hopfield model, although the
dynamical rules of the game differ from the Glauber dynamics
considered in neural networks \cite{KozlMars03}. The possibility of
retrieval states in majority games is discussed in \cite{DeMaGiarMose03}.

In analogy to work on dilute neural networks, in which not every
neuron is connected to every other neuron, we study the effects of
dilution on the phase behaviour of the MG in this paper. We restrict
ourselves to the case in which the average number of connections per
agent scales with the number of agents in such a way that it becomes
infinite in the thermodynamic limit as opposed to the case of finite
connectivity. In terms of the decision making of the agents this
corresponds to a game in which the agents do not use the global
aggregate bid to update the scores of their strategies, but only the
cumulative bid of an individual subset of neighbours to which they are
connected. Related games in which players adjust their decisions
dependent on the behaviour of agents in a local neighbourhood have
been discussed for example in \cite{PaczBassCorr00, MoelDeLo01,
KaliSchuBrie00}. MG-type models with inter-agent communication on
random graphs have also been studied analytically using a so-called
'Crowd-Anticrowd Theory' in \cite{GourChoeHuiJohn04, JohnHui03}.

\section{Model definitions}
Before introducing the dilution we will briefly recall the dynamical
rules of conventional MGs as studied for example in \cite{HeimCool01,
CoolHeim01}. The $N$ agents in the MG will be labelled with Roman
indices. At each round $t$ of the game each agent $i$ takes a trading
decision $b_i(t)\in{\rm I\!R}$ (a `bid') in response to the
observation of public information $I_{\mu(t)}$. While in the original
version of the MG the information coded for the actual market history
\cite{ChalZhan98,Cool04} we will here assume that $\mu(t)$ is chosen randomly
and independently from a set with $P=\alpha N$ possible values,
i.e. $\mu(t)\in\{1,\dots,\alpha N\}$. This is the so-called MG with
random external information. One then defines the rescaled total
market bid at round $t$ as $A(t)=N^{-1/2}\sum_i b_i(t)$. Each agent
$i$ holds a pool of $S\geq 2$ fixed trading strategies (look-up
tables) $\bR_{ia}=(R_{ia}^1,\dots,R_{ia}^P)$, with $a=1,\dots,S$.  If
agent $i$ decides to use strategy $a$ in round $t$ of the game, his or
her bid at this stage will be $b_i(t)=R_{ia}^{\mu(t)}$. All strategies
$\bR_{ia}$ are chosen randomly and independently before the start of
the game; they represent the quenched disorder of this problem. The
behaviour of the MG was found not to depend much on the value of $S$
\cite{ChalZhan98,ManuLiRiolSavi98}, nor on whether bids are discrete
or continuous \cite{CavaGarrGiarSher99}. For convenience, we choose
$S=2$ and $\bR_{ia}\in\{-1,1\}^P$ in this paper. The agents decide
which strategy to use based on points $p_{ia}(t)$ which they allocate
to each of their strategies. These virtual scores are based on their
success had they always played that particular strategy:
\begin{equation}
p_{ia}(t+1) = p_{ia}(t) - R_{ia}^{\mu(t)} A(t). \label{eq:scores}
\end{equation}
Strategies which would have produced a minority decision are thus
rewarded. At each round $t$ each player $i$ then uses the strategy in
his or her arsenal with the highest score, i.e. $b_i(t)=R_{i\tilde
a_i(t)}^{\mu(t)}$, where $\tilde a_i(t) = \mbox{arg max}_a\,
p_{ia}(t)$. For $S=2$ the rules (\ref{eq:scores}) can then be
simplified upon introducing the point differences
$q_i(t)=\frac{1}{2}[p_{i1}(t)-p_{i2}(t)]$. Thus agent $i$ plays
strategy $\bR_{i1}$ in round $t$ if $q_i(t)>0$, and $\bR_{i2}$ if
$q_i(t)<0$, so that his or her bid in round $t$ reads
$b_i(t)=\omega_i^{\mu(t)}+\sgn[q_i(t)]\xi_i^{\mu(t)}$, where
$\boldomega_i=\frac{1}{2}[\bR_{i1}+\bR_{i2}]$ and
$\boldxi_i=\frac{1}{2}[\bR_{i1}-\bR_{i2}]$. The above update rule for
the score differences $\{q_i\}$ can be written as
\begin{equation} \label{onlineupdate}
 q_i(t+1) = q_i(t) - \xi_i^{\mu(t)}\left[
\Omega^{\mu(t)}+\frac{1}{\sqrt{N}}\sum_j \xi_j^{\mu(t)}
\mbox{sgn}[q_j(t)]\right]
\end{equation}
with $\boldOmega=N^{-1/2}\sum_j \boldomega_j$. Equation
(\ref{onlineupdate}) defines the standard (or so-called `on-line')
MG. Alternatively, it is common in studies of the MG to define the
dynamics in terms of an average over all possible values of the
external information in (\ref{onlineupdate}). This corresponds to
updating the $\{q_i\}$ only every ${\cal O}(N)$ time-steps, and leads
to the the so-called `batch' MG \cite{HeimCool01}:

\be\label{eq:updatefullyconnected}
q_i(t+1)=q_i(t)-\frac{2}{N}\sum_j\left[\sum_{\mu=1}^P
\xi_i^\mu\xi_j^\mu s_j(t)+\sum_{\mu=1}^P \xi_i^\mu
\omega_j^\mu\right].  \ee Here $s_j(t)=\mbox{sgn}[q_j(t)]$.  We can
now introduce a dilute version of the batch MG by modifying the update
rules as follows: \be\label{eq:updatedilute}
q_i(t+1)=q_i(t)-\frac{2}{N}\sum_j\frac{c_{ij}}{c}\left[\sum_{\mu=1}^P
\xi_i^\mu\xi_j^\mu s_j(t)+\sum_{\mu=1}^P \xi_i^\mu
\omega_j^\mu\right]. \ee Here, the $c_{ij}\in\{0,1\}$
($i,j=1,\dots,N$) are additional quenched random variables. As before
the disorder corresponding to the strategy assignments is contained in
the variables $\{\xi_i^\mu, \omega_i^\mu\}$. $c$ is a new control
parameter and denotes the probability for a given link $c_{ij}$ to be
present, $P(c_{ij}=1)=c$. It hence controls the degree of dilution, on
average any given agent is connected to $cN$ other agents.  For $c=1$
the fully connected batch MG is recovered.  In the context of neural
networks the limit $\lim_{N\to\infty} c= 0$, while still $\lim_{N\to\infty}
cN=\infty$ is often referred to as the `extremely dilute` limit
\cite{Cool00a, Cool00b, Verb03}\footnote{Note that our definition of
$c$ differs from the one used in \cite{Cool00a, Cool00b} by a factor
of $N$. There $c$ denotes the average number of connections per
neuron, and hence the extremely dilute limit corresponds to
$\lim_{N\to\infty} cN^{-1}=0$ and $\lim_{N\to\infty} c=\infty$. The
notation of \cite{Cool00a, Cool00b} is more convenient if the case of
finite-connectivity is considered.}. We will specify the further
details of the distribution from which the $\{c_{ij}\}$ are drawn
below.

In terms of the updating of strategy scores the dilution means that a
given agent $i$ no longer uses the total re-scaled bid
$A(t)=N^{-1/2}\sum_j b_j(t)$ to evaluate the performance of his or her
strategies, but takes into account only the cumulative bid
$A_i(t)=N^{-1/2}\sum_{\{j:c_{ij}\neq 0\}} b_j(t)$ of the agents $j$ to
which he or she is connected, i.e. for which $c_{ij}=1$. Such
so-called `local' MGs and related models have been studied in
\cite{PaczBassCorr00, MoelDeLo01, GourChoeHuiJohn04, JohnHui03}. In
local MGs the perceived market history may be different for different
agents. Accordingly, if the information given to the agents is based
on the real-market history, different agents should potentially
receive different pieces of information. In the remainder of this
paper we will not focus on interpreting the dilute MG in terms of real
markets, but will restrict ourselves to the statistical mechanics
analysis of the model defined by the update rule
(\ref{eq:updatedilute}).

It remains to specify the distribution from which the dilution
variables $\{c_{ij}\}$ are drawn. We will choose $c_{ii}=1$, and will
assume that $c_{k\ell}$ is independent of $c_{ij}$ whenever
$(k,\ell)\not\in\{(i,j),(j,i)\}$.  The parameters characterising the
distribution of the dilution variables are then the expectation value
and covariance 
\be c=\bra c_{ij}\ket_c=\bra c_{ji}\ket_c, ~~~~~~ \bra
c_{ij}c_{ji}\ket_c-c^2=\Gamma c(1-c), \ee 
note that the $\{c_{ij}\}$ only take values $0$ and $1$ so that $\bra
c_{ij}^2\ket_c=c$. We write averages over the dilution variables as
$\bra \dots \ket_c$. The parameter $\Gamma$ controls the symmetry of
the dilution and allows for a smooth interpolation between the case of
fully symmetric dilution $c_{ij}=c_{ji}$ for $\Gamma=1$ and the fully
uncorrelated case $\Gamma=0$, where $c_{ij}$ and $c_{ji}$ are chosen
independently.  If $c=1$, then all the links are present and $\Gamma$
becomes meaningless. In closed form the corresponding distribution for any
pair $c_{ij},c_{ji}$ with $i<j$ can be written as
\BE\label{eq:cdistrib}
P(c_{ij},c_{ji})&=&c\left[c\left(1-\Gamma\right)+\Gamma\right]\delta_{c_{ij},1}\delta_{c_{ji},1}
\nonumber \\
&&+c(1-c)(1-\Gamma)\left(\delta_{c_{ij},1}\delta_{c_{ji},0}+\delta_{c_{ij},0}\delta_{c_{ji},1}\right)
\nonumber \\
&&+(1-c)\left[\left(1-c\right)\left(1-\Gamma\right)+\Gamma\right]\delta_{c_{ij},0}\delta_{c_{ji},0}.
\EE Taking into account that on average there are $cN$ connections per
agent, the appropriate control parameter is now $\alpha=P/(cN)$, where
$P$ is as before the number of possible values of the information, i.e.
we have $\mu\in\{1,\dots,\alpha c N\}$.
\section{Dynamics}
\subsection{Generating functional and single effective agent process}
We will start by studying the dynamics of the dilute batch minority
game. The now standard approach is based on dynamical functionals
\cite{HeimCool01,CoolHeim01} and
has been applied to different versions of the MG in \cite{HeimDeMa01, DeMa03,
DeMaGiarMose03, SherGall03, GallCoolSher03}. Before we
introduce the generating functional let us re-write the update rule
(\ref{eq:updatedilute}) as follows
\be
q_i(t+1)=q_i(t)-\sum_j J_{ij}^c s_j(t)-\sum_j h_{ij}^c+\theta_i(t),
\ee
where we have introduced the shorthands
\be
J_{ij}^c=\frac{c_{ij}}{c}\frac{2}{N}\sum_{\mu=1}^P \xi_i^\mu\xi_j^\mu, ~~~~~~~~~~~
h_{ij}^c=\frac{c_{ij}}{c}\frac{2}{N}\sum_{\mu=1}^P
\xi_i^\mu\omega_j^\mu,
\ee
and have added the perturbation fields $\{\theta_i(t)\}$ to generate
response functions.

The dynamical partition function is then introduced as
\BE \label{eq:genfct}
\hspace{-2cm}Z[\boldpsi]&=&\bra e^{i\sum_{it}s_i(t)\psi_i(t)}\ket_{paths}
\nonumber\\
\hspace{-2cm}
&=& \int Dq D\hat q p_0(q(0)) \exp\left(i\sum_{it}
\hat q_i(t)(q_i(t+1)-q_i(t)-\theta_i(t))\right)\nonumber \\ 
\hspace{-2cm}&&\times
\exp\left(i\sum_{it}\hat q_i(t)\left(\sum_j J_{ij}^c s_j(t)+\sum_j
h_{ij}^c\right)\right)\exp\left(i \sum_{it}
s_i(t)\psi_i(t)\right), 
\EE 
where $Dq D\hat q=\prod_{it}dq_i(t)d\hat q_i(t)/(2\pi)$, and with
$p_0(q(0))$ the distribution of the starting values $\{q_i(0)\}$. Once
the disorder-averaged generating functional $\overline{Z[\boldpsi]}$
has been calculated correlation and response functions can be computed
by taking suitable derivatives with respect to the source and
perturbation fields $\psi_i(t)$ and $\theta_i(t)$, respectively.

Carrying out the average over the quenched disorder in
(\ref{eq:genfct}) requires a lengthy, but standard calculation. One
first averages over the dilution variables $\{c_{ij}\}$ and then
proceeds along the lines of \cite{HeimCool01} in order to perform the
average over the $\{\xi_i^\mu\}$ and $\{\omega_i^\mu\}$, i.e. over the
strategies assigned at the beginning of the game. We will not report
the details of the calculation here, but refer to \cite{Cool00a,
Verb03} where similar averaging procedures are discussed in the
context of dilute neural networks.

As usual the evaluation of $\overline{Z[\boldpsi]}$ converts the
original system (\ref{eq:updatedilute}) of $N$ coupled Markovian
processes into an equivalent effective non-Markovian process for a
single representative agent.  After setting the source fields
$\psi_i(t)$ to zero and assuming that $\theta_i(t)=\theta(t)$ for all
$i$ we find the following single effective agent equation 

\BE q(t+1)&=&q(t)-\alpha (1-c)s(t) +\theta(t)-\alpha c \sum_{\tp}
(\id+G)^{-1}_{t\tp} s(\tp)\nonumber \\ &&+\Gamma \alpha
(1-c)\sum_{\tp} G_{t\tp}s(\tp)+\theta(t)
+\sqrt{\alpha}\eta(t)\label{eq:effagentdilute}, \EE where
$s(\tp)=\mbox{sgn}(q(\tp))$. $\eta(t)$ is Gaussian noise with zero
mean and temporal correlations given by \be\label{eq:noisecorr} \bra
\eta(t)\eta(\tp)\ket_* =
c\left((\id+G)^{-1}(E+C)(\id+G^T)^{-1}\right)_{t\tp}+(1-c)
(E+C)_{t\tp}.  \ee $\id$ denotes the identity matrix,
$\id_{t\tp}=\delta_{t\tp}$, while $E$ has all entries equal to one,
$E_{t\tp}=1$. The matrices $C$ and $G$ are the dynamical order
parameters of the problem and have to be determined self-consistently
according to \be\label{eq:selfconscg} C_{t\tp}=\bra s(t)s(\tp)\ket_*,
~~~~~~~ G_{t\tp}=\bra\frac{\partial s(t)}{\partial\theta(\tp)}\ket_*,
\ee where $\bra\dots\ket_*$ denotes averages over the single-agent
noise $\{\eta(t)\}$. This procedure is exact in the thermodynamic
limit $N\to\infty$ and the order parameters $C$ and $G$ can be
identified with the disorder-averaged correlation and response
functions of the original multi-agent batch process \BE
C_{t\tp}&=&\lim_{N\to\infty}N^{-1}\sum_i\overline{\bra
s_i(t)s_i(\tp)\ket}_{paths}\\
G_{t\tp}&=&\lim_{N\to\infty}N^{-1}\sum_i\frac{\partial}{\partial
\theta_i(\tp)}\overline{\bra s_i(t)\ket}_{paths}. \EE
Eqs. (\ref{eq:effagentdilute}, \ref{eq:noisecorr},
\ref{eq:selfconscg}) define a self-consistent problem to be solved for
the macroscopic observables $C$ and $G$. We note that the symmetry
parameter $\Gamma$ is irrelevant in the case of full connectivity,
$c=1$ and that the single-effective agent problem reduces to that of
the standard batch MG \cite{HeimCool01} in this limit. Because of the
retarded interaction-terms and the presence of coloured noise, a full
analytical solution is in general not feasible beyond the first few
time-steps. Alternatively one can resort to a Monte-Carlo integration
of the single agent-process using the procedure of
\cite{EissOppe92}. This method allows to determine the correlation and
response functions numerically without finite-size effects, but
becomes more and more costly as the number of time-steps is increased.

\subsection{Stationary state}
We now proceed to study the stationary states of the single-effective
agent process. We make the common assumptions of time-translation
invariance

\be \lim_{t\to\infty} C_{t+\tau,t}=C(\tau), ~~~~~
\lim_{t\to\infty} G_{t+\tau,t}=G(\tau), \ee finite integrated response
\be \lim_{t\to\infty}\sum_{\tp\leq t} G_{t\tp}=\chi<\infty \ee and
weak long-term memory \be \lim_{t\to\infty} G_{tt^\prime}=0 \quad
\forall \tp \mbox{ finite}.  \ee 

Under these assumptions we can follow
\cite{HeimCool01} and perform an average over time of
(\ref{eq:effagentdilute}). One defines
$s=\lim_{\tau\to\infty}\tau^{-1}\sum_{t\leq\tau} \sgn[q(t)]$ as well
as $\tilde q=\lim_{t\to\infty} q(t)/t$ and obtains after setting $\theta(t)=0$ 
\be
\tilde q = - R s + \sqrt{\alpha} \eta, 
\ee where we have defined \be
R=-\alpha(1-c)(\Gamma\chi-1)+\frac{\alpha c}{1+\chi}. \ee 
$\eta=\lim_{\tau\to\infty}\tau^{-1}\sum_{t\leq\tau} \eta(t)$ is a
Gaussian random variable with zero-mean and variance \be
Y=\bra\eta^2\ket_*=c\frac{1+Q}{(1+\chi)^2}+(1-c)(1+Q), \ee where we
have introduced the persistent correlation $Q=\lim_{\tau\to\infty}
\tau^{-1}\sum_{t\leq \tau} C(t)$.

Now, as usual in studies of MGs we can distinguish self-consistently
between so-called `frozen' and `fickle' agents. The representative
agent is frozen and keeps playing the same strategy if
$|q(t)|\to\infty$ asymptotically, so that $\tilde q\neq 0$. In this
case one has $s=\sgn[\tilde q]$. Provided that $R>0$ this is easily
seen to be the case if $|\eta|>R/\sqrt{\alpha}$. On the other hand the
effective agent is fickle and keeps switching between both of his or
her strategies when $\tilde q=0$, which is the case if
$|\eta|<R/\sqrt{\alpha}$. In this case
$s=(\sqrt{\alpha}/R)\eta$. Given $Q=\bra s^2\ket_*$, the persistent
part of the correlation function can then be computed
self-consistently from

\BE\label{eq:raweqc} Q&=&\bra
\theta\left(|\eta|-R/\sqrt{\alpha}\right)\ket_*+\bra
\theta\left(R/\sqrt{\alpha}-|\eta|\right)\frac{\alpha\eta^2}{R^2}\ket_*
\nonumber \\ &=& \phi+\bra
\theta\left(R/\sqrt{\alpha}-|\eta|\right)\frac{\alpha\eta^2}{R^2}\ket_*,
\EE 
where $\theta(t)$ is the step-function, $\theta(t)=1$ for $t>0$
and $\theta(t)=0$ otherwise.  $\phi=\bra
\theta\left(|\eta|-R/\sqrt{\alpha}\right)\ket_*$ denotes the fraction
of frozen agents.  On the other hand, only fickle agents contribute to
the response function asymptotically and we have 
\be\label{eq:raweqchi}
\chi=\frac{1}{\sqrt{\alpha}}\bra\theta\left(R/\sqrt{\alpha}-|\eta|\right)\frac{\partial
s}{\partial\eta}\ket_* = \frac{1-\phi}{R}.  
\ee
Explicitly performing the Gaussian integrals over $\eta$ in
(\ref{eq:raweqc}) then leads to the following two non-linear, but
closed equations for $Q$ and $\chi$:

\BE
Q&=&1+\erf\left(\frac{R}{\sqrt{2\alpha Y}}\right)\left(\frac{\alpha
Y}{R^2}-1\right) -\sqrt{\frac{2}{\pi}}\frac{\sqrt{\alpha
Y}}{R}\exp\left(-\frac{R^2}{2\alpha Y}\right), \label{eq:dilsc1}\\
\chi&=&\frac{1}{R}\erf\left(\frac{R}{\sqrt{2\alpha Y}}\right).\label{eq:dilsc2}
\EE
The fraction of frozen agents reads $\phi=1-\erf\left(\frac{R}{\sqrt{2\alpha Y}}\right)$.

\begin{figure}[t]
\vspace*{1mm}
\begin{tabular}{cc}
\epsfxsize=72mm  \epsffile{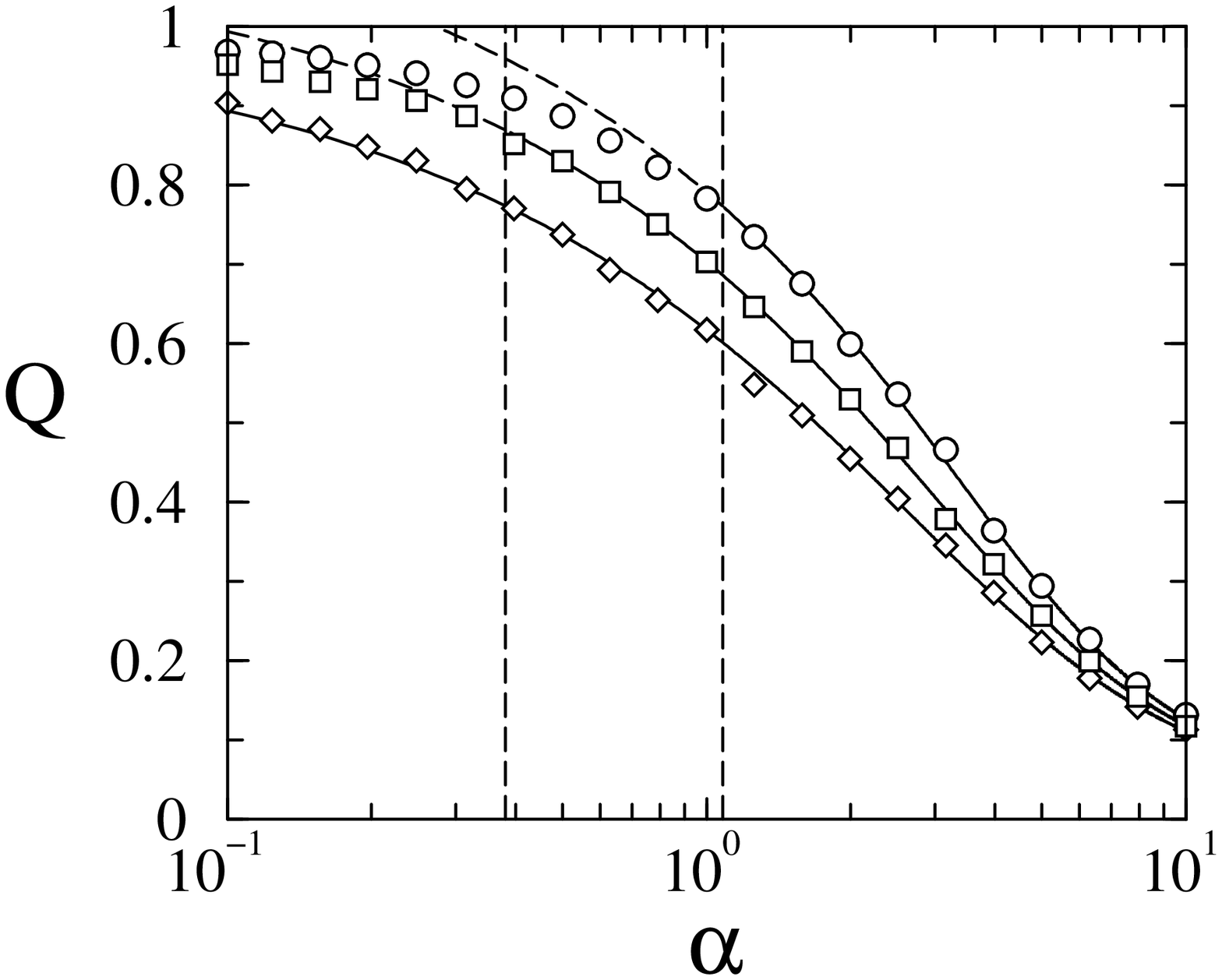} ~&~~
\epsfxsize=72mm  \epsffile{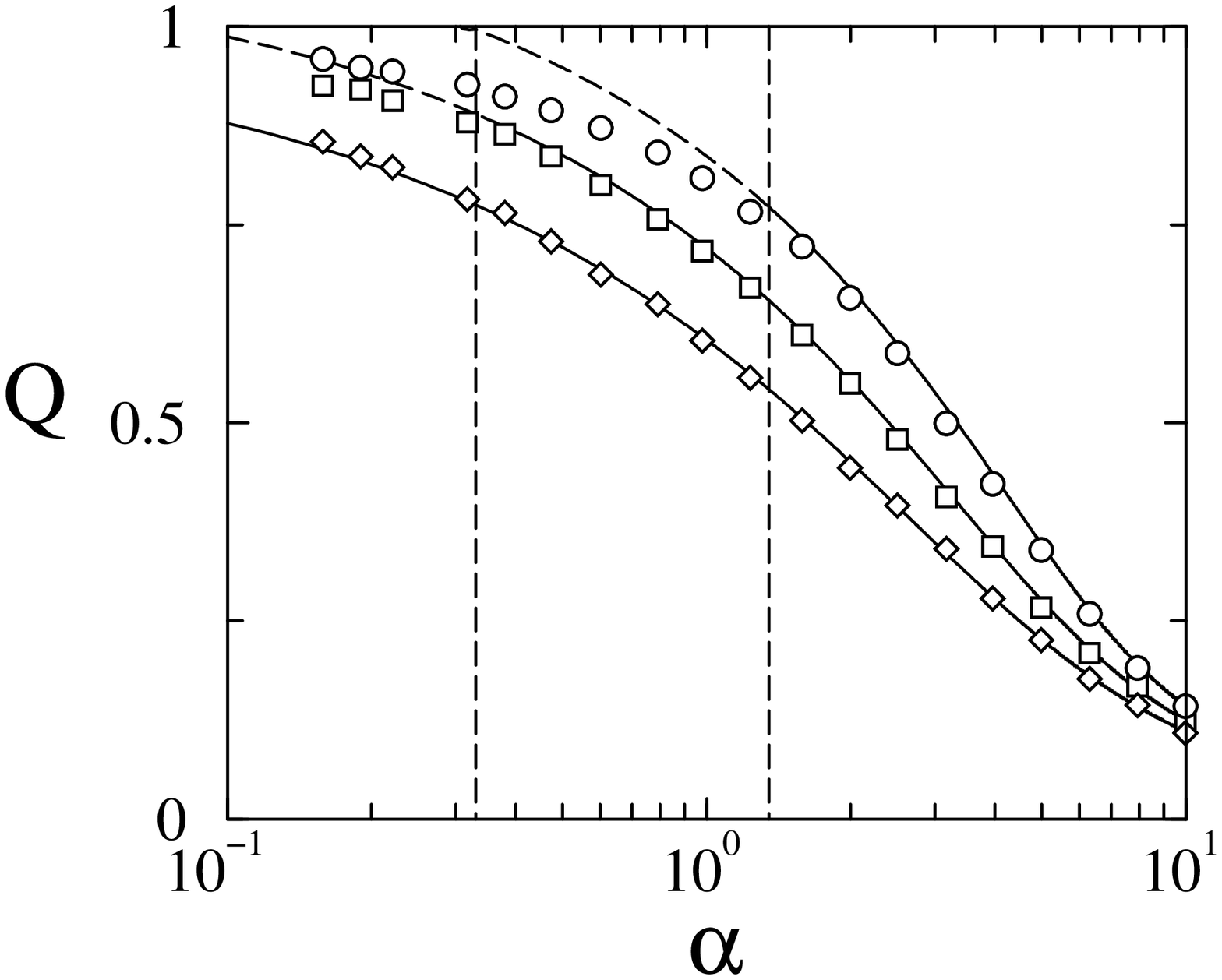}
\end{tabular}
\vspace*{4mm} \caption{Persistent correlation $Q$ as a function of
$\alpha$. Markers are simulations for $\Gamma=1$ (circles),
$\Gamma=0.5$ (squares) and $\Gamma=0$ (diamonds). The solid lines are
the theoretical predictions for the phase with weak long-term memory, they
have been continued as dashed lines below the transition, where they
are no longer valid. Vertical dashed lines mark the onset of memory for
$\Gamma=0.5$ and $\Gamma=1$, respectively. No
transition is predicted for $\Gamma=0$. Left:
$c=0.5$. Simulations are performed for $N=500$ players, measurements
are taken over $250$ time-steps preceded by $250$ equilibration steps.
The results are averaged over $10$ realisations of the disorder. 
Right: Limit of extreme dilution, $c=0$ (simulations are for $N=1000$
players with $c=N^{-1/2}$, $20$ disorder samples). }
\label{fig:theory_q_c}
\end{figure}

The system (\ref{eq:dilsc1}, \ref{eq:dilsc2}) reduces to the
corresponding equations of the standard batch MG \cite{HeimCool01} in
the limit $c=1$. It can be solved numerically for any fixed values of
the parameters $0\leq c \leq 1$ and $0\leq \Gamma \leq 1$ to  obtain
the order parameters $Q$ and $\chi$ as a function of $\alpha$. We
compare the analytical results for the persistent part of the
correlation function with numerical simulations in
Fig. \ref{fig:theory_q_c}. Here, the
extremely dilute limit $\lim_{N\to\infty} c=0$, $ \lim_{N\to\infty}
cN=\infty$ has been simulated by setting $c=N^{-1/2}$. We find near
perfect agreement for large values of $\alpha$, but for $\Gamma>0$
systematic deviations are observed in the low-$\alpha$ region.  Using
the method of \cite{EissOppe92} we have performed an explicit
iteration of the single-agent equation to confirm that the deviations
for small values of $\alpha$ are not artifacts due to a finite number
of agents in the simulations.

In the standard MG, $c=1$, the breakdown of the ergodic theory is
associated with the onset of anomalous response, i.e. the violation of
the assumption that the integrated response $\chi$ be finite. At the
same time one expects that the assumption of weak long-term memory is
broken as well \cite{DeMaThesis}. No singularities are found, however,
in the numerical solution of (\ref{eq:dilsc1}, \ref{eq:dilsc2}) for
$c<1$. Since no signs of ageing are found in MGs, we are lead to the
conclusion that the assumption of weak long-term memory breaks down
below a critical value of $\alpha$, whereas $\chi$ remains finite. We
will refer to this as the onset of memory, similar behaviour was
observed in batch MGs with market-impact correction
\cite{HeimDeMa01}. 

\subsection{Onset of memory and phase diagram} 
We will now proceed to identify this onset of memory analytically.  To
this end we follow the lines of \cite{HeimDeMa01} and separate
time-translation invariant (TTI) contributions to the response
function from the non-TTI parts and write \be \lim_{t\to\infty}
G_{t\tp}=\tilde G(t-\tp)+\hat G_{t\tp}.  \ee While perturbations in
the initial stages of the dynamics might influence the final state of a
given agent, we expect that the effect becomes independent of $t$
eventually, so that we assume $\lim_{t\to\infty} \hat G_{t\tp}=\hat
G(\tp)$. In the stationary state, the point differences of frozen
agents have become large so that only fickle agents are affected by
perturbations applied at later times. This is accounted for by $\tilde
G$. Hence we expect that $\lim_{\tp\to\infty} \hat G(\tp)=0$
\cite{HeimDeMa01}. Assuming furthermore that $\hat G$ is small we
expand the single-effective agent process to first oder in $\hat G$
and find

\BE
\hspace{-2cm}q(t+1)&=&q(t)-\alpha (1-c)s(t) -\alpha c \sum_{\tp}
(\id+\tilde G)^{-1}_{t\tp} s(\tp)+\Gamma \alpha (1-c)\sum_{\tp}
\tilde G_{t\tp}s(\tp)\nonumber \\ 
\hspace{-2cm}&&+\theta(t)+\sqrt{\alpha}\eta(t) +\alpha c
\sum_{\tp} \sum_{n=0}^\infty\sum_{m=0}^{n-1} \left[(-\tilde G)^m\hat G
(-\tilde G)^{n-m-1}\right]_{t\tp} s(\tp)\nonumber \\ 
\hspace{-2cm}&&+\Gamma\alpha
(1-c) \sum_{\tp} \hat G(\tp) s(\tp)+{\cal O}(\hat G^2).  \EE

Writing $\chi = \sum_t \tilde G(t)$  one then has after averaging over time 

\BE 
\hspace{-2cm}
\tilde q &=&
-\alpha(1-c) s -\frac{\alpha cs}{1+\chi}+\Gamma\alpha(1-c)\chi s
+\sqrt{\alpha}\eta \nonumber\\ 
\hspace{-2cm}
&&+\alpha c
\sum_{n=0}^\infty\sum_{m=0}^{n-1} (-\tilde \chi)^m\sum_{\tp}\hat
G(\tp)\sum_{\td} (-\tilde G)^{n-m-1}_{\tp\td} s(\td)
+\Gamma\alpha(1-c)\sum_{\tp}\hat G(\tp)s(\tp).  \EE 

As before we
have $\chi=\frac{1}{\sqrt{\alpha}}\bra\frac{\partial
s}{\partial\eta}\ket_*$, whereas 

\BE  
\hat G(t)&=&\bra \frac{\partial
s}{\partial\theta(t)}\ket_*\nonumber\\
&=&\gamma^{-1}\left[\sqrt{\alpha} c \sum_{n=0}^\infty\sum_{m=0}^{n-1}
(-\chi)^m\sum_{\tp}\hat G(\tp)\sum_{\td} (-\tilde
G)^{n-m-1}_{\tp\td} \tilde G_{\td t}\right.\nonumber \\
&&\left.+\Gamma\sqrt{\alpha}(1-c)\sum_{\tp}\hat G(\tp)\tilde G_{\tp t}
\right], \EE where we have introduced 
$\gamma=\sqrt{\alpha}(1-c)(1-\Gamma\chi)+\sqrt{\alpha}c/(1+\chi)$.

We now define $\hat\chi = \sum_t \hat G(t)$, and after summing over $t$
and re-organising the terms we finally find to first order in $\hat G$
\be \hat \chi =
\sqrt{\alpha}\frac{\chi}{\gamma}\left[\frac{c}{(1+\chi)^2}+\Gamma(1-c)\right]\hat\chi+{\cal
O}(\hat G^2). \ee 

Although $\hat\chi=0$ is always a solution, a bifurcation occurs when
the following condition is fulfilled \be\label{eq:moline}
\sqrt{\alpha}\chi\left[\frac{c}{(1+\chi)^2}+\Gamma(1-c)\right]=\gamma,
\ee so that non-zero solutions for $\hat\chi$ become possible. For
fixed $\Gamma$ this condition marks the continuous onset of long-term
memory and defines a line in the $(\alpha,c)$-plane. We will refer to
this line as the memory-onset (MO) line in the following, and will
discuss the resulting phase diagram in section \ref{sec:pg}.

An intuitive argument for the different type of phase transition of
the dilute MG as compared with the fully connected model can be based
on a simple geometrical explanation of the transition in the fully
connected standard MG \cite{MarsChalZecc00,HeimelThesis}. In the
stationary state only the $(1-\phi)N$ fickle agents contribute to the
integrated response $\chi$, as perturbations do not affect frozen
agents. Now from Eqs. (\ref{onlineupdate}) and
(\ref{eq:updatefullyconnected}) one realizes that the $N$-dimensional
vector $\qbo=(q_1,\dots,q_N)$ moves in the space spanned by the
$P=\alpha N$ vectors $\boldxi^\mu=(\xi_1^\mu,\dots,\xi_N^\mu)$. If
$P>(1-\phi)N$ this space is certain to contain the homogeneous
perturbation $\theta_i=\theta$, and any small perturbation in this
direction will be washed out over time. If, however, $P<(1-\phi)N$
then the whole space is no longer spanned by the $\{\boldxi^\mu\}$ and
some perturbations can not be removed so that the integrated response
diverges at the value $\alpha_c$ of $\alpha$, at which
$\alpha_c=1-\phi(\alpha_c)$, which is exactly what is found in the
fully connected MG
\cite{HeimCool01, MarsChalZecc00, ChalMarsZhan00}. In the dilute case
(\ref{eq:updatedilute}) the vector $\qbo$ no longer moves in the space
spanned by the $\{\boldxi^\mu\}$, as the agents use local bids
$A_i(t)$ to update their strategy scores. From this argument it is
plausible that the phase transition marked by an onset of diverging
integrated response is no longer found as soon as $c<1$. 

\section{Replica calculation and AT-instability}
\subsection{General replica theory}
We will now turn to the analysis of the statics of the dilute version
of the MG. By considering a continuous time limit of the update rules
of the standard MG is it possible to show that the stationary state in
the ergodic regime $\alpha>\alpha_c$ corresponds to minima of the disordered function 
$H_1=(1/P)\sum_{\mu=1}^P\left[\sum_{i,j}\xi_i^\mu\xi_j^\mu\phi_i\phi_j+2\sum_{i,j}\xi_i^\mu\omega_j^\mu\phi_i+\sum_{i,j}\omega_i^\mu\omega_j^\mu\right]$ \cite{ChalMarsZecc00, MarsChalZecc00}. 
The variables $\phi_i\in[-1,1]$ are the average asymptotic values of
$s_i(t)$ so that the corresponding agent is frozen whenever
$\phi_i=\pm 1$ and remains fickle for $-1<\phi_i<1$.

We generalise the above function to the case of dilution
and introduce \be
H_c=\frac{1}{2\alpha}\left[\sum_{ij}\left\{J_{ij}^c\phi_i\phi_j+2h_{ij}^c\phi_i+K_{ij}^c\right\}\right].
\ee 
Here we use the same definitions for $J_{ij}^c$ and $h_{ij}^c$ as in
the analysis of the dynamics, and introduce new variabes
$K_{ij}^c=\frac{c_{ij}}{c}\frac{2}{N}\sum_{\mu=1}^P
\omega_i^\mu\omega_j^\mu$. 

As usual in disordered systems it is in general not possible to find a
Hamiltonian which is minimised by the dynamics in the case of
asymmetric couplings. Therefore, we have to restrict this
section to the case of fully symmetric dilution, $c_{ij}=c_{ji}$,
i.e. we will assume $\Gamma=1$ in the above distribution
(\ref{eq:cdistrib}) of the connectivity variables.

In order to minimise the above function $H$ in terms of the variables
$\{\phi_i\}$, we resort to the replica method and compute the
disorder-average of the replicated partition function \be
Z^n=\int_{-1}^1 \left(\prod_{ia}d\phi_i^a\right)\exp\left[-\frac{\beta}{2\alpha}\sum_a\sum_{ij}\left\{J_{ij}^c\phi_i^c
\phi_j^a+2h_{ij}^c\phi_i^a+K_{ij}^c\right\}\right]  \ee 
at some `annealing temperature' $\beta^{-1}$. We will eventually take
the limit $\beta\to\infty$. The superscript $a$ is a replica index,
$a=1,\dots,n$. The minima of $H_c$ are then found as \be
\mbox{min}_{\{\phi_i\}}
H_c=-\lim_{\beta\to\infty}\lim_{N\to\infty}\lim_{n\to 0}
\frac{\overline{Z^n}-1}{\beta N n}.  \ee The computation of
$\overline{Z^n}$ is again lengthy, but straightforward and is
essentially a combination of the corresponding calculations for the
fully connected MG \cite{ChalMarsZecc00, MarsChalZecc00, ChalMarsZhan00} and replica
analyses of dilute neural networks \cite{Cool00b,Verb03}. We will
therefore only report the final result:
\be\label{dilutereplicapresaddlepoint} \overline{Z^n}=\int DQ DS
\exp\left(-\beta n N f(Q,S)+{\cal O}(N^0)+{\cal O}(n^2)\right),  \ee
with $DQ=\prod_{ab}[\sqrt{N}dQ_{ab}/\sqrt{2\pi}]$ and similarly for
$DS$. Here the replicated `free energy' is given by
\BE\label{dilutefreeenergy}
\hspace{-2cm}f(Q,S)&=&\frac{\alpha
c}{2\beta n}\log\det T + \frac{\alpha \beta}{2n}\sum_{ab}S_{ab}Q_{ab}
-\frac{1}{\beta n}\log\left[\int_{-1}^1 \left(\prod_{a}d\phi^a\right)
\exp\left(\frac{\alpha\beta^2}{2}\sum_{ab}S_{ab}\phi^a\phi^b\right)\right]\nonumber
\\ \hspace{-2cm}&&-\frac{\beta}{4\alpha
n}(1-c)\sum_{ab}Q_{ab}Q_{ab}-\frac{c-1}{2n}\sum_a
Q_{aa}-\frac{\beta}{\alpha}\frac{(1-c)}{2n}\sum_{ab}Q_{ab}+\frac{1-c}{2}.
\EE 
$Q$ is the overlap-matrix defined by $Q_{ab}=N^{-1}\sum_i \phi_i^a
\phi_i^b$, and the $S_{ab}$ are the conjugate Lagrange multipliers. 
The $n\times n$-matrix $T$ is given by
$T_{ab}=\delta_{ab}+\frac{\beta}{\alpha}(1+Q_{ab})$.
\subsection{Replica symmetric ansatz}
We will now make a replica symmetric ansatz corresponding to the
assumption of ergodicity in the dynamics and will set \be
Q_{ab}=q+(Q-q)\delta_{ab}, ~~~~~~S_{ab}=s+(S-s)\delta_{ab}.  \ee

Insertion into (\ref{dilutefreeenergy}) and taking the limit $n\to 0$
gives the replica-symmetric free-energy 
\BE\label{diluteRSf}
\hspace{-2cm}f_{RS}(Q,S)&=&\frac{\alpha c}{2\beta}\log\left
[ 1+\frac{\beta}{\alpha}(Q-q)\right]+\frac{\alpha
c}{2}\frac{(1+q)}{\alpha+\beta(Q-q)}\\ 
\hspace{-2cm}&&+\frac{\alpha
\beta}{2}(SQ-sq)-\frac{1}{\beta} \bra\left[\log\int_{-1}^1 d\phi
\exp\left(-\beta V_z(\phi)\right)\right]\ket_z\nonumber \\ 
\hspace{-2cm}&&
-\frac{\beta}{4\alpha}(1-c)(Q^2-q^2)-\frac{c-1}{2}Q+\frac{\beta}{2\alpha}(1-c)(q-Q)+\frac{1-c}{2},
\EE 
where $V_z(\phi)$ is an effective potential defined by
$V_z(\phi)=-\sqrt{\alpha s}z\phi-\frac{\alpha\beta}{2}(S-s)\phi^2$. $\bra\dots\ket_z$ denotes the average over the
standard Gaussian variable $z$.

The integrals in (\ref{dilutereplicapresaddlepoint}) are then
performed by the method of steepest descent in the thermodynamic limit
and the corresponding saddle-point equations are obtained by working
out the variations of $f_{RS}$ with respect to the parameters $Q,q, S$
and $s$. Similar to \cite{ChalMarsZhan00} one finds
\BE s&=&c\frac{1+q
}{(\alpha+\beta(Q-q))^2}+\frac{1-c}{\alpha^2}(1+q) \\ Q&=& \bra \bra
\phi^2|z\ket_\phi\ket_z\\
\beta(S-s)&=&-\frac{c}{\alpha+\beta(Q-q)}+\frac{\beta(1-c)}{\alpha^2}(Q-q)+\frac{c-1}{\alpha}\\
\beta(Q-q)&=&\frac{1}{\sqrt{\alpha s}} \bra z \bra \phi|z\ket_\phi\ket_z.
\EE

In these expressions we have introduced the average \be \bra f(\phi)
|z\ket_\phi = \frac{\int_{-1}^1 d\phi f(\phi)\exp\left(-\beta
V_z(\phi)\right)}{\int_{-1}^1 d\phi \exp\left(-\beta
V_z(\phi)\right)}.  
\ee 
We will be looking for solutions with
$\lim_{\beta\to\infty} Q=\lim_{\beta\to\infty} q$
and $\lim_{\beta\to\infty} S=\lim_{\beta\to\infty} s$ and will also use the quantities 
\be
\chi=\frac{\beta}{\alpha}(Q-q),\qquad
\zeta=-\sqrt{\frac{\alpha}{s}}\beta(S-s), 
\ee
which will remain finite in this limit. The remaining averages
over $\phi$ are easily evaluated in the limit $\beta\to\infty$ as the
corresponding integrals over $\phi$ are dominated by the minimum of
the potential $V_z(\phi)$ in $\phi\in[-1,1]$. We find that the minimum
is at $\phi=1$ for $z\geq\zeta$ and at $\phi=-1$ for
$z\leq-\zeta$. This is the analogue of the corresponding condition on
the time-averaged single-particle noise for frozen agents in the
generating-functional calculation. For $-\zeta<z<\zeta$ the minimum is
found at $\phi=z/\zeta$, so that the agent is fickle. Working out the
remaining integrals we then find the following two equations for the
order parameters $\chi$ and $Q$ \cite{ChalMarsZhan00}: 
\BE Q &=&
1-\sqrt{\frac{2}{\pi}}\frac{\exp(-\zeta^2/2)}{\zeta}-\left(1-\frac{1}{\zeta^2}\right)\erf\left(\frac{\zeta}{\sqrt{2}}\right),\\
\sqrt{\alpha
s}\beta(Q-q)&=&\frac{1}{\zeta}\erf\left(\frac{\zeta}{\sqrt{2}}\right)
.  \EE 
In order to make contact with the results obtained from the
analysis of the dynamics we define 
\BE
R&=&-\alpha^2\beta(S-s)=-\alpha(1-c)(\chi-1)+\frac{\alpha c}{1+\chi}
\\ Y&=&\alpha^2 s = c \frac{1+Q}{(1+\chi)^2}+(1-c)(1+Q) \EE so that
$\zeta=R/\sqrt{\alpha Y}$ and find \BE Q &=&
1-\sqrt{\frac{2}{\pi}}\frac{\sqrt{\alpha
Y}}{R}\exp\left(-\frac{R^2}{2\alpha Y}\right)-\left(1-\frac{\alpha
Y}{R^2}\right)\erf\left(\frac{R}{\sqrt{2\alpha
Y}}\right),\label{eq:replspeq1}\\
\chi&=&\frac{1}{R}\erf\left(\frac{R}{\sqrt{2\alpha
Y}}\right)\label{eq:replspeq2} .  
\EE 

\begin{figure}[t]
\vspace*{1mm}
\begin{tabular}{cc}
\epsfxsize=72mm  \epsffile{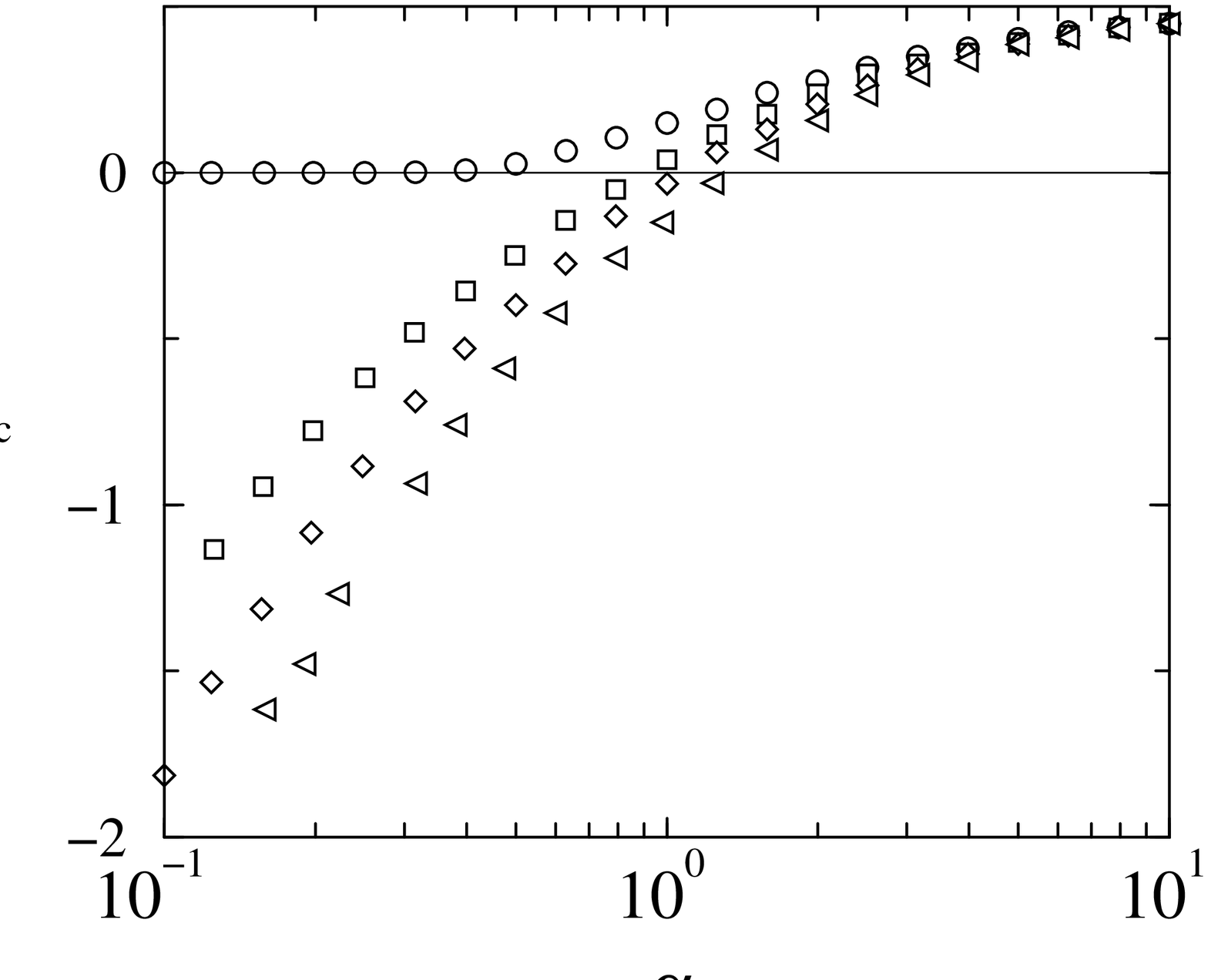} ~&~~
\epsfxsize=72mm  \epsffile{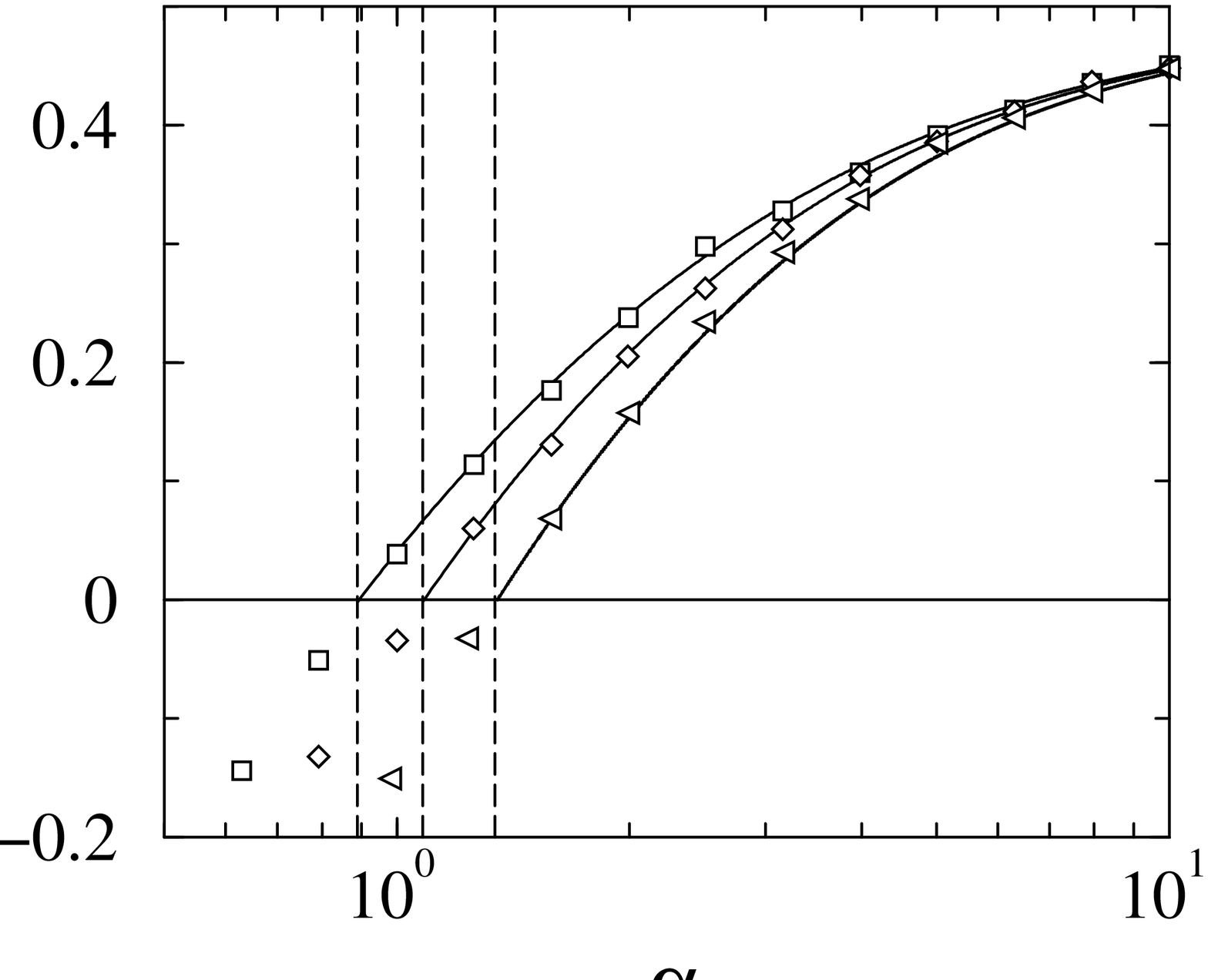}
\end{tabular}
\vspace*{4mm} \caption{Left: stationary value of the `energy' $H_c$ as a
function of $\alpha$ for $c=1$ (circles), $c=0.75$ (squares), $c=0.5$
(diamonds) and $c=0$ (triangles). Data for $c>0$ are from simulations with
$N=500$ agents, run for $500$ time-steps and averaged over $20$
realisations of the disorder. Data for $c=0$ is for $N=1000$ and $10$
realisations. Right: zoom for large values of $\alpha$, the solid
lines are the predictions of the replica-symmetric theory,
Eq. (\ref{eq:finalRSf}), the vertical dashed lines mark the onset
of the AT-instability, see Eq. (\ref{eq:atline})} \label{fig:energy}
\end{figure}

\begin{figure}
\vspace*{1mm}
\begin{center}
\epsfxsize=85mm  \epsffile{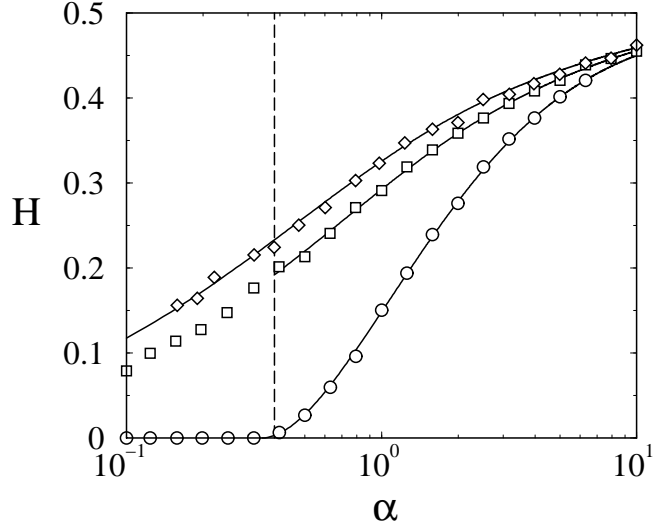}\\
\end{center}
\vspace*{4mm} \caption{Predictability $H$ as a function of $\alpha$ for $(c=0,\Gamma=0)$, $(c=0.5, \Gamma=0.5)$ and the fully connected case $c=1$ (from top to bottom). Open symbols are data from numerical simulations of the batch process, solid lines are the theoretical predictions of Eq. (\ref{eq:predict}) in the ergodic phase. The vertical dashed line marks the onset of memory for $(c=0.5, \Gamma=0.5)$. No memory onset occurs for $(c=0, \Gamma=0)$. (Simulation parameters are $N=500$ agents with data averaged over $50$ realisations of the disorder and $500$ time-steps for $c=1$ and $c=0.5$, $N=1000$ agents and $200$ realisations for $c=0$.)}
\label{fig:predictability}
\end{figure}

Note that this set of equations is identical to the one found from the
dynamics (cf. (\ref{eq:dilsc1},\ref{eq:dilsc2})) for the case of fully
symmetric dilution, $\Gamma=1$. Finally, the replica symmetric free
energy (\ref{diluteRSf}) can be simplified upon using the saddle-point
equations and we find: \be\label{eq:finalRSf}
\lim_{\beta\to\infty}f_{RS}=\frac{c}{2}\frac{1+Q}{(1+\chi)^2}+\frac{1-c}{2}\left(1+Q\right)\left(1-2\chi\right).
\ee  We compare
this prediction of the replica-symmetric theory with simulations in
Fig. \ref{fig:energy} and find very good agreement in the phase with
only weak-long term memory. For $c=1$ the result of
Eq. (\ref{eq:finalRSf}) reduces to the replica-symmetric free
energy of the fully connected model as reported in
\cite{ChalMarsZhan00}. Note that in this case $H_{c=1}$ can be
identified as the so-called predictability of the market and is a
positive definite quadratic form and hence non-negative. A similar interpretation of $H_c$
is not straightforward in the dilute model, and in particular $H_c$
need not be non-negative for $c<1$. It is however instructive to study the predictability 
\be
H=\lim_{N\to\infty}\frac{1}{P}\sum_{\mu=1}^P \bra A|\mu\ket^2=\lim_{N\to\infty}\frac{1}{PN}\sum_{\mu=1}^N\sum_{ij} (\omega_i^\mu+\xi_i^\mu\phi_i) (\omega_j^\mu+\xi_j^\mu\phi_j)
\ee
also for the case $c<1$. Here $\bra A|\mu\ket$ denotes an average over
the total bid conditioned on a pattern $\mu$ in the stationary state \cite{ChalMars99, ChalMarsZecc00}.
We display data from simulations in Fig. \ref{fig:predictability}. An
analytical result for the predictability in the ergodic state of the
fully connected batch MG is given in \cite{HeimCool01} and reads
\be\label{eq:predict}
H=\frac{1}{2}\frac{1+Q}{(1+\chi)^2}.
\ee
We note that this expression can carried over to the dilute case
$c<1$ and that the numerical data agrees very well with the prediction of (\ref{eq:predict}), see Fig. \ref{fig:predictability}\footnote{In the fully connected case the same expression for the
predictability $H=H_{c=1}$ can also be obtained from the replica
analysis of the statics, as mentioned in the main text. Note, however,
that the result of \cite{HeimCool01} is obtained from a generating
functional analysis of the dynamics of the fully connected
game. Eq. (\ref{eq:predict}) therefore applies not only to the case of
symmetric interactions between the agents, but also for arbitrary
symmetry parameters $\Gamma<1$ (where no replica theory is available due to the lack of symmetry in the couplings $\{J_{ij}^c\}$).}.

While the two phases in the fully connected model correspond to
$H_{c=1}=0$ in the regime of low $\alpha$ and to a predictable phase
($H_{c=1}>0$) above $\alpha_c(c=1)=0.3374\dots$, we observe that no
unpredictable (or so-called symmetric) phase is present for dilution
parameters $c<1$. Although Fig. \ref{fig:predictability} only depicts
results for some combinations of the parameters $c$ and $\Gamma$, we
have verified that $H$ is strictly positive also for other values of
$\Gamma$ as long as $c<1$. This absence of a symmetric phase with
vanishing predictability is also observed in MGs with market impact
correction \cite{damienthesis,MarsChalZecc00}.

\subsection{Calculation of the AT-line}

We will now study the stability of the ground states of $H_c$ obtained
within the replica-symmetric approximation against small fluctuations
which break replica-symmetry. To this end we follow the procedure
first proposed by de Almeida and Thouless \cite{deAlThou78} and study
the eigenvectors of the matrix of second derivatives of the free
energy with respect to the $\{Q_{ab}\}$ and $\{S_{ab}\}$. This results
in an instability condition marking the onset of replica-symmetry
breaking and defines a line in the $(\alpha,c)$-plane, the so-called
AT-line. Our analysis follows the lines of \cite{DeMaMars01,
DeMaThesis}, where the AT-line has been computed for the fully
connected MG with market impact correction.

The stability matrix of second derivatives of the free energy has
dimension $n(n-1)\times n(n-1)$ is given by \be
\Sigma=\left(\begin{tabular}{cc} A & C \\ C & B \end{tabular}\right),
\ee where the sub-matrices $A, B$ and $C$ read \be
A_{(ab)(cd)}=\frac{\partial^2(nf)}{\partial Q_{ab}\partial Q_{cd}},
~~~~ B_{(ab)(cd)}=\frac{\partial^2(nf)}{\partial S_{ab}\partial
S_{cd}}, ~~~~ C_{(ab)(cd)}=\frac{\partial^2(nf)}{\partial
Q_{ab}\partial S_{cd}}.  \ee The derivatives with respect to the
$\{Q_{ab}\}$ are found as \cite{DeMaMars01, DeMaThesis} \BE
A_{(ab)(ab)}&=&
-c\frac{\beta}{\alpha}\left[C_1^2+C_2^2\right]-\frac{\beta}{\alpha}(1-c)
\\ A_{(ab)(ac)}&=&-c\frac{\beta}{\alpha}C_1\left[C_1+C_2\right]\\
A_{(ab)(cd)}&=&-c\frac{2\beta}{\alpha}C_1^2, \EE where
$C_1=-\frac{\beta(1+Q)}{\alpha(1+\chi)^2}$ and
$C_2=C_1+\frac{1}{1+\chi}$.  Furthermore we find \BE
B_{(ab)(ab)}&=&-\alpha^2\beta^3\left[\bra\bra\phi^2|z\ket_\phi^2\ket_z-\bra\bra\phi|z\ket_\phi^2\ket_z^2\right]\\
B_{(ab)(ac)}&=&-\alpha^2\beta^3\left[\bra\bra\phi^2|z\ket_\phi
\bra\phi|z\ket_\phi^2\ket_z-\bra\bra\phi|z\ket_\phi^2\ket_z^2\right]\\
B_{(ab)(cd)}&=&-\alpha^2\beta^3\left[\bra\bra\phi|z\ket_\phi^4\ket_z
-\bra \bra\phi|z\ket_\phi^2\ket_z^2\right] \EE and finally one has \be
C_{(ab)(cd)}=\alpha\beta\delta_{ac}\delta_{bd}.
\ee

We now study fluctuations $\delta Q_{ab}$ and $\delta S_{ab}$ around
the replica symmetric solution, which are of the form $\delta
S_{ab}=x\delta Q_{ab}$, with $x$ a proportionality constant
\cite{DeMaMars01}.  We will assume that the perturbation is symmetric,
i.e $\delta Q_{ab}=\delta Q_{ba}$ and $\delta S_{ab}=\delta S_{ba}$
and that it does not affect the diagonal elements so that $\delta
Q_{aa}=\delta S_{aa}=0$.  We will here consider fluctuations of the
form \BE \delta Q_{12}&=&\frac{1}{2}(3-n)(2-n)y \\ \delta
Q_{1b}=\delta Q_{2b}&=&
\frac{1}{2}(3-n)y ~~~ b\notin\{1,2\} \\ \delta
Q_{ab}&=&y~~~~~~~~~~~~~~ a,b\notin\{1,2\}, \EE corresponding to the
so-called `replicon' mode \cite{deAlThou78, dotsenko}.

In the limit $n\to 0$ the corresponding eigenvalues are found to be \cite{DeMaMars01,DeMaThesis}
\be
\lambda_{\pm}=-\frac{1}{2}(u+v)\pm\frac{1}{2}\sqrt{(u-v)^2+4},
\ee where \be u=\frac{c}{\alpha^2(1+\chi)^2}+\frac{1-c}{\alpha^2}, ~~~
v=\alpha\beta^2\bra
\left(\bra\phi^2|z\ket_\phi-\bra\phi|z\ket_\phi^2\right)^2\ket_z.  \ee
Now, $\lambda_-$ never changes sign, so that the instability sets in
when $\lambda_+=0$. This happens when $uv=1$, i.e. when \be
\lim_{\beta\to\infty}\beta^2\bra
\left(\bra\phi^2|z\ket_\phi-\bra\phi|z\ket_\phi^2\right)^2\ket_z
=\left(\frac{c}{\alpha(1+\chi)^2}+\frac{1-c}{\alpha}\right)^{-1}.  \ee
It remains to evaluate the averages on left hand side. This is
conveniently done by introducing
$F_z(h)=\beta^{-1}\log\int_{-1}^1d\phi e^{-\beta
V_z(\phi)+h\beta\phi}$ and upon taking derivatives with respect to $h$
\cite{DeMaThesis}. After some algebra we finally find the following
condition for the AT-line
\be\label{eq:atline}
\chi\left(\frac{c}{(1+\chi)^2}+1-c\right)=\frac{c}{1+\chi}+(1-c)(1-\chi).
\ee This coincides with the MO-line obtained from the dynamics
(\ref{eq:moline}), provided we set $\Gamma=1$ in the dynamical
calculation. A comparison of Eq. (\ref{eq:finalRSf}) and
(\ref{eq:atline}) demonstrates that the minima of $H_c$ obtained from
the replica-symmetric theory become unstable precisely at the
zero-crossing of $H_c$, as illustrated in Fig. \ref{fig:energy}.

\section{Phase diagram and discussion}\label{sec:pg}
\begin{figure}
\vspace*{1mm}
\begin{center}
\epsfxsize=85mm  \epsffile{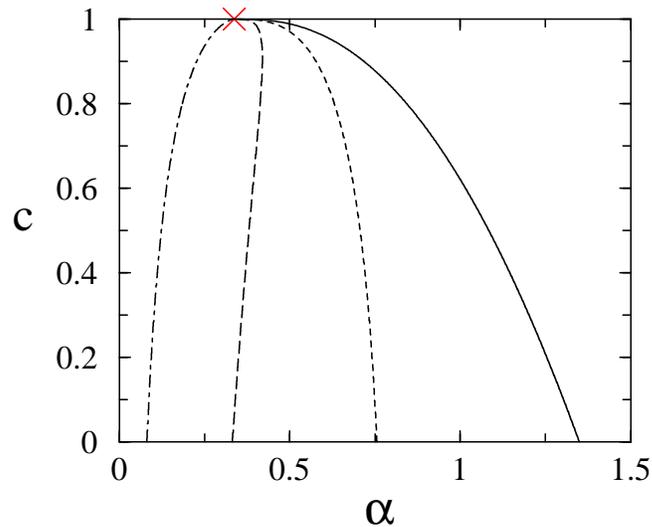}\\
\end{center}
\vspace*{4mm} \caption{Phase diagram in the $(\alpha,c)$-plane. The
lines separate phases with memory at low-$\alpha$ from phases without
long-term memory and indicate the location of the transition for
$\Gamma=0.25$ (dot-dashed line), $\Gamma=0.5$ (dashed), $\Gamma=0.75$
(dotted) and $\Gamma=1.0$ (solid line). The solid line
($\Gamma=1$) coincides with the AT-line obtained from the replica
calculation. The cross marks the critical point of the fully
connected MG at $\alpha_c=0.3374\dots$.}
\label{fig:phasediagram_dilute}
\end{figure}

The resulting phase diagram is shown in
Fig. \ref{fig:phasediagram_dilute}. We report the transition lines in
the $(\alpha,c)$-plane for different fixed values of $\Gamma>0$.  They
separate an ergodic phase with weak long-term memory at high values of
$\alpha$ from a phase with memory for low $\alpha$. For $\Gamma=1$ the
replica symmetric ground state of $H_c$ is stable to the right of the
MO(=AT)-line. Assuming that $\chi$ remains finite we find that the
condition for the MO-line (\ref{eq:moline}) is given by
$1-c+c/(1+\chi)^2=0$ for $\Gamma=0$ and that hence no onset of memory
occurs in the case of fully asymmetric dilution as the left-hand side
is always positive. We find that the values of $\alpha_c(c,\Gamma)$
approach the critical point $\alpha_c=0.3374\dots$ of the fully
connected MG continuously in the limit $c\to 1$. Note that the
parameter $\Gamma$ is meaningless in the fully connected case $c=1$.

\begin{figure}
\vspace*{1mm}
\begin{center}
\epsfxsize=85mm  \epsffile{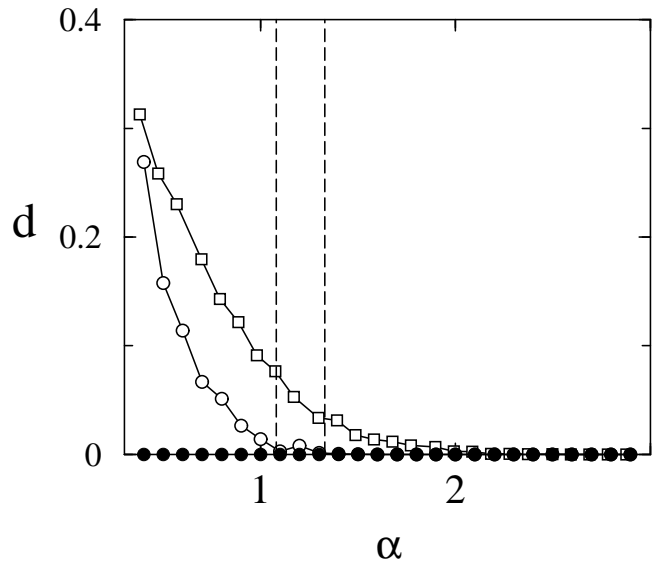}\\
\end{center}
\vspace*{4mm} \caption{Average distance $d$ between the stationary
states of two identical copies of the system as a function of
$\alpha$. All runs are started from zero initial conditions. Open
markers are obtained by applying a small perturbation at $t=0$ (circles:
$c=0.5$, squares: $c=0$; $\Gamma=1$ in both cases). The vertical
dashed lines indicate the location of the onset of memory,
Eq. (\ref{eq:moline}). The full markers are obtained by perturbing the
system at a later time $t=500$.  All simulations are run up to
$500$ batch steps after the perturbation is applied. $10$ runs
with different random perturbations are generated per disorder sample
and the average is taken over all pairwise distances. Results are then
averaged over at least $20$ realisations of the disorder. }
\label{fig:distance}
\end{figure}

To give further support for the picture of memory-onset we follow
\cite{HeimDeMa01} and measure the distance $d=N^{-1}\sum_i
(\phi_i-\phi_i^\prime)^2$ between two microscopic stationary states obtained
from simulations of two copies of the system for an identical
realisation of the disorder, but starting from slightly different
initial conditions. In this expression $\phi_i$ and $\phi_i^\prime$ are the
long time averages of $\sgn[q_i]$ and $\sgn[q_i^\prime]$ in the
stationary state, respectively. In Fig. \ref{fig:distance} we report
the average distance $d$ as a function of $\alpha$ for different values of the
parameters $(c,\Gamma)$. For large $\alpha$ the two copies end up in
the same microscopic state so that $d=0$ indicating that the long-term
memory is weak. Below the predicted onset of memory the assumption of
weak long-term memory is broken, and we find $d>0$ so that the final
microscopic stationary state depends on the initial conditions. While
the data reported in Fig. \ref{fig:distance} qualitatively supports
the picture of a continuous onset of memory, a
verification of the precise point of memory-onset by measuring $d$ in
simulations turns out to be difficult due to finite-size
effects. Finally, we find that the two copies end up in the same
state if the perturbation is applied at much later times, so that we
indeed conclude that $\lim_{\tp\to\infty} \hat G(\tp)=0$.

\begin{figure}[t]
\vspace*{1mm}
\begin{tabular}{cc}
\epsfxsize=72mm  \epsffile{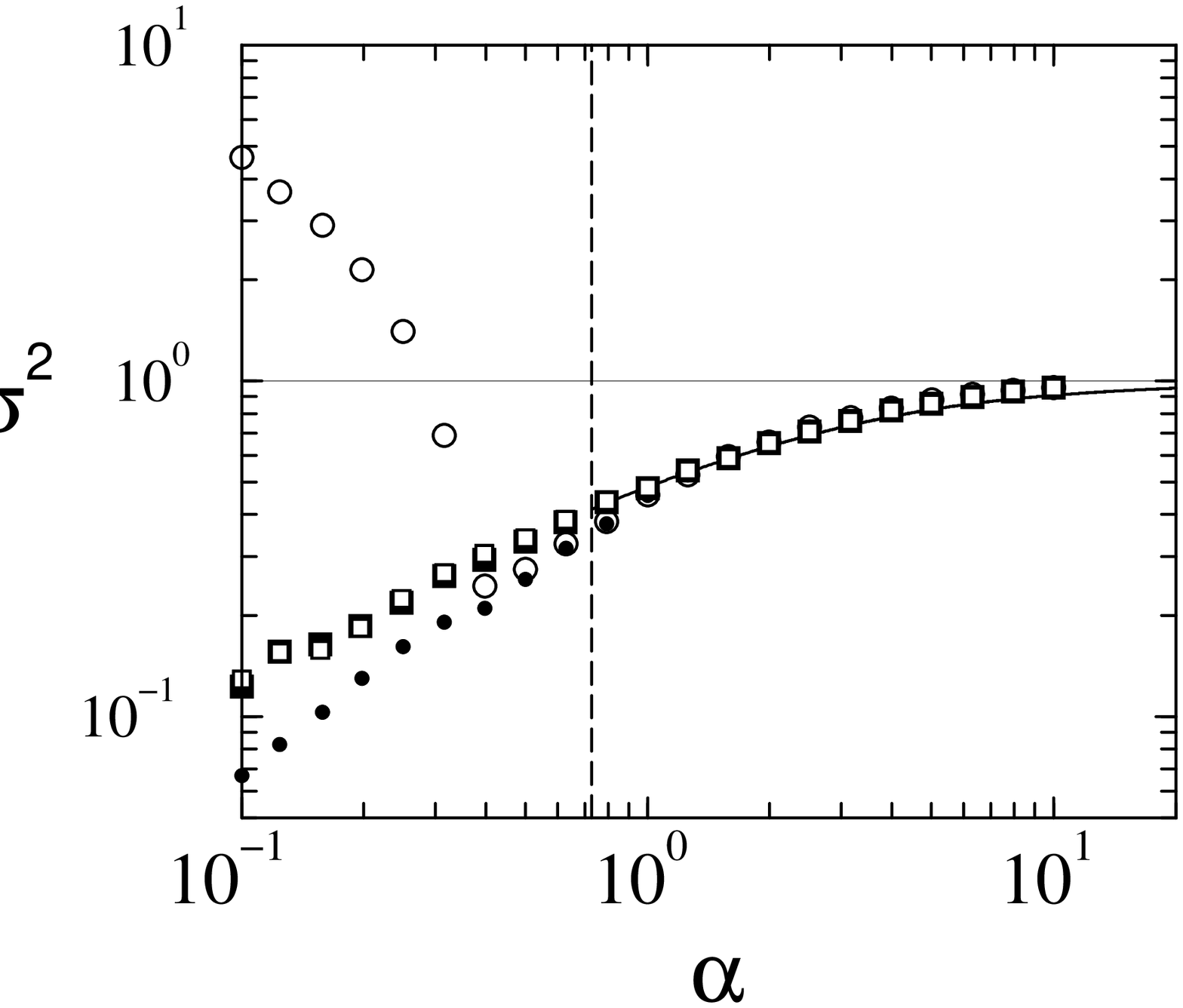} ~&~~
\epsfxsize=72mm  \epsffile{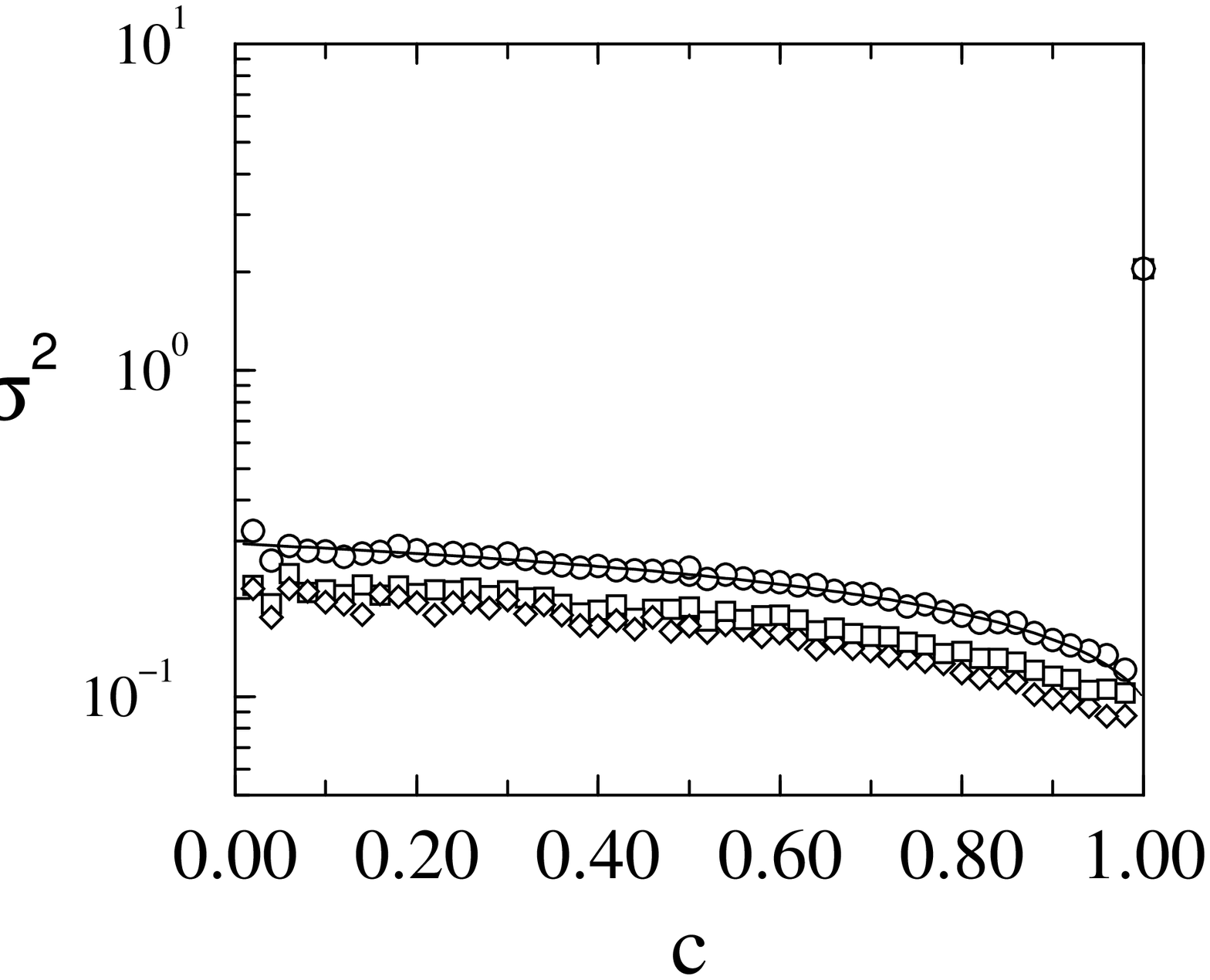}
\end{tabular}
\vspace*{4mm} \caption{Left: Volatility $\sigma^2$ vs $\alpha$
for $c=0.25$, $\Gamma=0.75$ (squares). Markers represent numerical
simulations with $N=1000$ players, averaged over $20$ samples of the
disorder. Open symbols are for {\em tabula rasa} starts, $q_i(0)=0$,
full markers for biased starts, $|q_i(0)|=10$.  The vertical dashed
line marks the breakdown of the ergodic theory for $c=0.25,
\Gamma=0.75$. The solid line to the right is the approximation of
Eq. (\ref{eq:volapprox}) for $c=0.25, \Gamma=0.75$. The circles are
for comparison and show the results of the standard batch MG,
$c=1$. $\sigma^2=1$ is the random trading limit. Right: $\sigma^2$ vs
$c$ at fixed $\alpha=0.2$ for $\Gamma=0$ (circles), $\Gamma=0.5$
(squares)and $\Gamma=1.0$ (diamonds), simulations started from zero
initial conditions. The solid line is the approximation of
(\ref{eq:volapprox}) for the case $\Gamma=0$, in which the long-term
memory is weak at $\alpha=0.2$, so that the ergodic theory
applies. For $\Gamma=0.5$ and $\Gamma=1$ the fixed value $\alpha=0.2$
is below the onset of memory.\label{fig:vol}}
\end{figure}

We conclude this section by reporting results for the magnitude of the fluctuations of
the total bid $A(t)$ in Fig. \ref{fig:vol}. $\sigma^2=\bra
A^2\ket_{time}$ measures the global efficiency of the market;
the better supply and demand match, the smaller $\sigma^2$ and the more
efficient the markt. By construction $<A>_{time}=0$. Note that
in our model $\sigma^2$ measures the fluctuations of the total bid and
not of the local bids $A_i$ perceived by the individual agents. The
volatility in the stationary state is found as \cite{HeimCool01} 

\be
\sigma^2=\frac{1}{2}\lim_{t\to\infty}\left[(\id+G)^{-1}(E+C)(\id+G^T)^{-1}\right]_{tt}.
\ee
Following the lines of \cite{HeimCool01} it is possible to find an
approximate expression for the volatility in terms of the persistent
order parameters $\phi$ and $\chi$, which holds in the ergodic phase:
\be\label{eq:volapprox}
\sigma^2=\frac{1+\phi}{2(1+\chi)^2}+\frac{1}{2}(1-\phi).
\ee
As shown in Fig. \ref{fig:vol} this approximation is in very good
agreement with numerical simulations of the batch process.

While the stationary volatility in the fully connected MG started from
zero initial conditions exhibits a minimum at $\alpha_c=0.33\dots$, we
find an increasing function $\sigma^2(\alpha)$ for diluted games, see
left panel of Fig. \ref{fig:vol}. Although only the case $c=0.25,
\Gamma=0.75$ is reported in Fig. \ref{fig:vol}, we find very similar
curves for other values of $c<1$ and $\Gamma$. Analogous behaviour of
$\sigma^2$ was reported for games with market impact correction
\cite{MarsChalZecc00}. We observe that despite the memory-effects no
significant dependence of the volatility on the initial conditions is
found in dilute games. This is in sharp contrast with the standard MG,
where one finds stationary states in which volatility $\sigma^2$
diverges as $\alpha\to0$ for so-called {\em tabula rasa} starts,
$q_i(0)=0$ for all $i$ and low-volatility solutions for strongly
biased starts, $|q_i(0)|={\cal O}(1)$. In the dilute game, only
low-volatility states appear to be realised.  The right panel of
Fig. \ref{fig:vol} demonstrates that the volatility is discontinuous
as $c\uparrow 1$ for fixed $\alpha<\alpha_c(c=1)=0.3374\dots$. In
fact, depending on $\alpha$ the jump can extend to several orders of
magnitude. Again, very similar behaviour has been reported in MGs with
market-impact correction \cite{MarsChalZecc00}.

Finally, we have verified in numerical simulations that the behaviour
of the model does not change much, if the batch update rule
(\ref{eq:updatedilute}) is replaced by a dilute version of the on-line
MG, with random external information (see Eq. (\ref{onlineupdate}) for
the definition of the fully connected on-line MG). We do not report
the numerical data here, but would only like to remark that order
parameters in the ergodic stationary state such as the persistent
correlation $Q$ follow the same curves as the corresponding
observables of the batch game (as displayed in
Fig. \ref{fig:theory_q_c}), and that the phase diagram does not appear
to be sensitive to the choice of on-line or batch learning rules. The
volatilities of the dilute on-line and batch games differ slightly in
their quantitative values, but their qualitative behaviour as a
function of $\alpha$, $c$ and $\Gamma$ coincides. In particular, for
on-line games, we find only low volatility solutions for low
$\alpha$ and $c<1$, and a jump in the volatility as $c\uparrow 1$ for
small enough fixed values of $\alpha$, similar to the data displayed
for the batch game in Fig. \ref{fig:vol}b). These findings are
consistent with the results of \cite{CoolHeim01}, where it was
observed that order parameters such as $Q$ and $\phi$ in the
stationary state are identical in the fully connected on-line and
batch MGs, and that only slight quantitative differences are present
in their respective volatilities.
 
\section{Conclusions}
In summary, we have presented static and dynamical analyses of the
dilute batch MG with random external information. Using a generating
functional approach we have first solved the dynamics of the
game. Assuming time-translation invariance, finite integrated response
and weak long-term memory we have computed the persistent order
parameters in the stationary state.  Numerical simulations confirm
these findings convincingly. Standard methods can be employed to
obtain a satisfactory approximation for the magnitude of the global
market fluctuations of the model.

The study of the dynamics is complemented by the replica analysis of
the statics of the game with fully symmetric dilution. We find that
the resulting equations for the order parameters agree with the ones
obtained from the dynamics.

Similar to studies of MGs with market impact correction a condition
for the breakdown of weak long-term memory can be derived from the
dynamics, and a continuous onset of memory is found at finite
integrated response. The resulting MO-line agrees with the AT-line
obtained from the statics, marking the onset of instability of the
replica-symmetric theory. While the microscopic stationary state of
the present model below the transitions depends on the state from
which the dynamics is started, macroscopic observables such as the
volatility appear to be mostly insensitive to initial conditions. This
indicates that below the transition many microscopic stationary states
exist, but all with the same or nearly the same global order
parameters. Simulations of the MG with market impact correction reveal
similar behaviour. In both models, the dilute MG presented in this
paper and the model with market impact correction, the microscopic
update rules are such that the players use individual bids $A_i(t)$ to update
their strategy scores, as opposed to the common global bid $A(t)$ in
the fully connected model without impact correction. Based on the
above geometrical considerations one might speculate that other
variants of the MG with this property might also exhibit replica symmetry
breaking and the onset of long-term memory at finite integrated
response.

The dilute model shares two other features with the MG with
market-impact correction: the discontinuity in $\sigma^2$ as the
special case of the conventional game is approached in its non-ergodic
phase, and the absence of a symmetric phase with vanishing
predictability.

We believe that the dilute model is an intrinsicially interesting
model from the point of view of statistical mechanics and hope that
the present contribution might serve as a starting point for further
analytical studies. For example an approximate or exact solution might
be attempted in the regime of broken weak long-term memory, but with
finite integrated response. It might also be interesting to apply the
recently developed methods to study the dynamics of disordered systems
with finite connectivity \cite{Hatchetal04, WemmCool03} to MGs
with a finite number of connections per agent and to understand the
consequences for the phase diagram.

While the present model is designed to be accessible by the tools of
equilibrium and non-equilibrium statistical mechanics and was not
primarily devised as a realistic model of a market, there is currently
much interest in the interplay between local connectivity and global
competition in networks of agents \cite{JohnHui03,
GourChoeHuiJohn04}. The work on the dilute MG presented in this paper
should therefore be seen as an intermediate step towards a better
understanding of more elaborate models. Further developments of the
current analytical theory might for example include dynamically
evolving networks, inter-agent communication or the introduction of
real local histories.

\section*{Acknowledgements}

The author would like to thank EPSRC for financial support under
research grant GR/M04426 and studentship 00309273. He also
acknowledges the award of a Rhodes Scholarship and support by Balliol
College, Oxford and by the European Community's Human Potential
Programme under contract HPRN-CT-2002-00319, STIPCO. Fruitful
discussions with D Challet, ACC Coolen, M Marsili and D Sherrington
are gratefully acknowledged.

\end{document}